\documentclass[reprint,nofootinbib,amsmath,amssymb,fontenc aps]{revtex4-2}

\usepackage{graphicx,dcolumn,bm,hyperref,float}
\usepackage{anyfontsize}
\usepackage{xcolor}

\newcommand{\mpl}{M_\text{\tiny{Pl}}}

\begin{document}
	
	\date{\today}
	
	\title{String axiverse enhancement of superradiant dark matter production}
	\author{Diogo~S.~Gorgulho$^{1,2,3}$}
	\email{d.severino.gorgulho@rug.nl}
	\author{Jacob~A.~Litterer$^{3,4}$}
	\email{jacob.litterer@fc.up.pt}
	\author{Jo{\~a}o~G.~Rosa$^{2,3}$}
	\email{jgrosa@uc.pt}
	\affiliation{
		$^1$Van Swinderen Institute for Particle Physics and Gravity, University of Groningen, Nijenborgh 3, 9747 AG Groningen, The Netherlands \\
		$^2$Univ Coimbra, Faculdade de Ci\^encias e Tecnologia da Universidade de Coimbra, Rua Larga, 3004-516 Coimbra, Portugal \\
		$^3$Centro de F\'isica da Universidade de Coimbra, Rua Larga, 3004-516 Coimbra, Portugal \\
		$^4$Departamento de F\'isica e Astronomia, Faculdade de Ci\^encias, Universidade do Porto, Rua do Campo Alegre s/n, 4169-007, Porto,
		Portugal 
		\looseness=-1}

	\begin{abstract}
		We study the effects of string axion emission on dark matter production by light primordial black holes (PBHs), through both evaporation and superradiance. We show, in particular, that the Hawking emission of $\mathcal{O}(100-10^5)$ light axion species predicted in realistic string theory constructions can significantly enhance the efficiency of superradiance, given the associated increase in the PBH spin. The string axiverse thus significantly expands the parametric regions (dark matter mass and PBH mass and spin) for which a sizeable fraction of dark matter may presently be in the form of ``micro-boson stars'': the self-gravitating remnants of superradiant dark matter clouds. Conversely, for too large a number of axion species PBHs evaporate too quickly for superradiant clouds to attain their maximum mass. Finally, assuming that all dark matter is produced by PBHs (through both superradiance and Hawking emission), we show that the axions emitted during PBH evaporation give an immeasurably small contribution to the relativistic degrees of freedom at recombination.
	\end{abstract}
	
	\maketitle

	\section{Introduction}\label{sec:intro}
	
	It is well established that most gravitating matter in the universe does not have significant interactions with visible sector particles, with a variety of possible explanations of this dark matter from fundamental physics. One compelling candidate would be a particle with mass around the weak scale, which would naturally freeze out at the right time to produce the observed relic dark matter density. With no (in)direct detection of such particles at $\mathcal{O}(1)$ GeV--TeV scales, we are motivated to consider scenarios for (a) heavier dark matter and (b) smaller couplings to the visible sector. In this work, we study a purely gravitational production mechanism of a heavy dark matter boson, which we assume has no direct coupling to the visible sector. 
	
	In this mechanism, dark matter is produced by the combination of evaporation and superradiance of small black holes in the early universe, with the black holes evaporating to produce all (or a fraction) of the dark matter before big bang nucleosynthesis (BBN). Primordial black holes (PBHs) themselves remain an interesting candidate for some or all of the dark matter \cite{Carr:2020gox}, but in this work we take the PBHs to be light enough that they evaporate before BBN; their role in this scenario is to end up with no relic density themselves, but instead produce the dark matter boson with the correct relic density.
	
	This scenario was originally studied in \cite{Bernal:2022oha,March-Russell:2022zll} (see also \cite{Jia:2025vqn}), which identified the parametric regions (natal PBH mass and spin and dark matter boson mass) where superradiant dark matter production is efficient compared to Hawking emission alone, which was considered in earlier works \cite{Lennon:2017tqq,Allahverdi:2017sks,Hooper:2019gtx,Bernal:2020bjf,Bernal:2020ili,Masina:2020xhk,Baldes:2020nuv,Gondolo:2020uqv,Cheek:2021odj,Cheek:2021cfe}.
	Superradiance produces the heavy dark matter boson more efficiently for a spin-1 candidate than for spin-0, and for larger black hole spin. Since typical PBHs are not expected to form with large spins \cite{Chiba:2017rvs,Mirbabayi:2019uph,DeLuca:2019buf} (though other possibilities abound \cite{Harada:2017fjm,Flores:2021tmc,Saito:2023fpt,Saito:2024hlj}), and given the upper bound on the mass of PBHs that evaporate before BBN, in realistic scenarios superradiance can only produce a small fraction of dark matter.
	
	Superradiant dark matter production is nevertheless interesting from the phenomenological perspective, since dark matter particles are produced in non-relativistic bound states, or ``clouds'', around the black holes. Reference \cite{March-Russell:2022zll} conjectured that these clouds could survive as (microscopic) self-gravitating boson stars after the PBHs evaporate. This is supported by numerical simulations of a (spherical, non-relativistic) scalar cloud around an evaporating black hole, for which \cite{Neves:2025kxp} showed that a large fraction of the cloud remains gravitationally self-bound if the black hole's evaporation occurs adiabatically until its mass falls below roughly 50\% of the cloud's mass. Even if these ``dark boson stars'' account for only a small fraction of the present dark matter abundance, this may dramatically change the way we should search for dark matter: while individual dark matter particles may be nearly impossible to detect if they have no interactions stronger than gravity, dark boson stars may be detectable as a result of their large occupation numbers and coherent enhancement of scattering cross sections (e.g.~\cite{Hardy:2015boa}).
	
	This motivates revisiting superradiant dark matter production in the context of the so-called ``string axiverse'' \cite{Fallon:2025lvn,Svrcek:2006yi,Arvanitaki:2009fg,Arvanitaki:2010sy,Gendler:2023kjt}, i.e.~the fairly general prediction of a large number of very light pseudoscalars in the low energy particle spectrum in string compactifications. One of these axion-like particles can explain the strong CP problem via the Peccei-Quinn mechanism; similarly, most of the axions are expected to resist perturbative corrections. We refer to all of these axion-like particles as ``axions'' for the remainder of this paper. 
	
	As found in \cite{Calza:2021czr,Calza:2023rjt}, a black hole emitting hundreds or even thousands of axion species through Hawking radiation in addition to the Standard Model particles (and the graviton) spins up as it evaporates. This may thus make superradiant dark matter production more efficient, and we will show that a substantial fraction of dark matter may result from PBH superradiance in this context for a broader range of black hole and dark matter masses, even at low black hole spins.
	
	While there are other scenarios in which one or more axions themselves play the role of dark matter \cite{Marsh:2015xka,OHare:2024nmr}, since the many axions of the string axiverse are typically very light, we will assume that they are not efficiently produced by superradiance but only through Hawking emission. The axions thus behave as dark radiation rather than dark matter, although we will show that they give a negligible contribution to the total radiation energy density in the parametric regimes relevant for superradiant dark matter production.	
	
	In this work, we make no attempt to explain the formation of PBHs, the precise number of axion species, or their masses. We assume there exist some number of light axion species (essentially massless), and that the PBHs form with a monochromatic mass spectrum and the number density required to produce the observed dark matter abundance. We also assume PBH evaporation follows Hawking's semi-classical treatment, without leaving a Planck-scale remnant \cite{Chen:2014jwq} (although this would not change our results). We take the dark matter to be a heavy spin-0 particle, a complex scalar field with only (minimal) gravitational coupling and no self-interactions (which we note could influence the dynamics of superradiance \cite{Baryakhtar:2020gao,Branco:2023frw,Xie:2025npy}). The dark boson carries a global charge whose conservation is necessarily violated by superradiance. This implies, in particular, that superradiant clouds and their boson star remnants (as opposed to axion clouds) are stable against gravitational wave emission and that any other processes potentially leading to their decay are Planck-suppressed, making their lifetimes naturally longer than the age of the universe.

	\section{Hawking emission and superradiance}\label{sec:theory0}
	
	In this section we briefly review the dynamics of black hole evaporation including superradiance and recap important previous results upon which we will build. More detailed treatments of these topics can be found in \cite{Hawking:1975vcx,Hartle:1976tp,Starobinskii:1973vzb,Cardoso:2004nk,Pani:2012bp,Rosa:2012uz,Pani:2012vp,Witek:2012tr,Brito:2014wla,Arvanitaki:2014wva,Brito:2015oca,Rosa:2017ury,Cardoso:2018tly,Baumann:2018vus,Baumann:2019eav}. 
	In the following, we set $c=\hbar=k_\text{B}=1$, so that all units are set by the Planck mass, $\mpl = 1.2\times 10^{19}$ GeV.
	
	\subsection{Black hole evaporation}
	
	A black hole described by the Kerr metric has dimensionless spin parameter $a \equiv \mpl^2 J/M^2$ and temperature 
	\begin{align}
		T_\text{H} = {\mpl^2\over 4\pi M} { \sqrt{1-a^2} \over \left( 1 + \sqrt{1-a^2} \right)} \label{TH}
	\end{align} 
	and emits all particles with mass $\mu<T_\text{H}$. These emitted particles carry away mass and angular momentum from the black hole,
	\begin{align}
		{d M \over d t} = - e_T {\mpl^4 \over M^2} ~~,~~~~{d J \over d t} = -e_J { J \mpl^4 \over M^3} \label{evapeqs}
	\end{align}
	where $(e_T,\,e_J)$ are the energy- and spin-emissivity coefficients, which count all accessible degrees of freedom $n_s$ with spin $s$
	\begin{align}
		e_T = \sum_s n_s f_s ~~,~~~~ e_J = \sum_s n_s g_s ~.
	\end{align}
	The functions $(f_s,\,g_s)$ come from adding up the contributions from all field modes for each spin,
	\begin{align}
		\begin{pmatrix}
			f_s \\
			g_s
		\end{pmatrix} = \sum_{\ell,\,m}  \int_0^\infty {d\left(\omega M\right) \over 2\pi} {\Gamma_{s,\ell m}\!\left(M,a,\omega\right) \over e^{\left(\omega-m\Omega\right)/T_\text{H}} \pm 1} 
		\begin{pmatrix}
			\omega M \\
			m_{} a^{-1}
		\end{pmatrix}
	\end{align}
	where the plus (minus) corresponds to fermions (bosons). The sums are carried out over quantum numbers $(\ell,\,m)$ of a spin-$s$ particle with energy $\omega$. The black hole has rotation speed $\Omega = a/2r_+$ (at the outer horizon $r_+$). The graybody factors $\Gamma_{s,\ell m}\!\left(M,a,\omega\right)$ reflect the suppressed emission of low energy particles, as they must tunnel through a potential barrier outside the event horizon to reach large distances. We treat all emitted particles as massless, since the PBHs evaporating before BBN have Hawking temperatures $T_H\sim 10 \, (M/10^6\ \mathrm{kg})^{-1}$ TeV.
	
\begin{figure}
		\centering
		\includegraphics[width=\columnwidth]{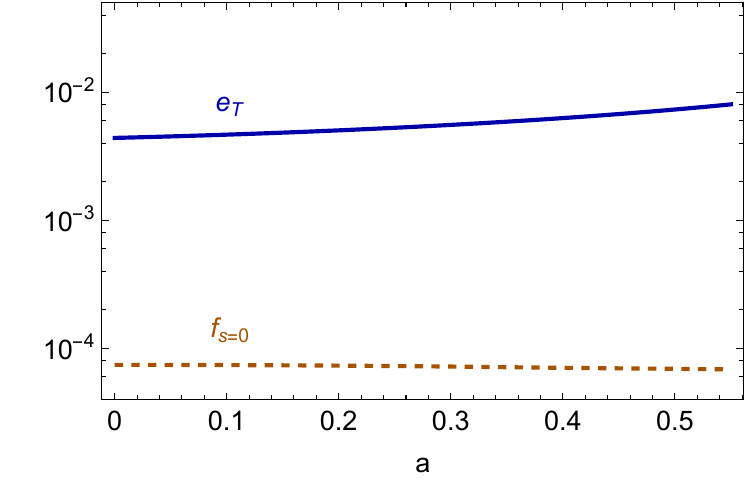}
		\caption{Interpolating functions of $f_0$ and the full $e_T$ including all SM particles and gravitons constructed from $f_s(a)$ values computed in \cite{Calza:2023rjt}. 
			%{\color{red}{\\(Keep figure but with $a$ not $a_0$, although I think this function is only a good approximation for spins up to 0.5-0.6)}}
		}
		\label{fig:eTandf0}
	\end{figure}

	The $(f_s,\,g_s)$ functions have been computed numerically in discrete $a$, from which we construct interpolating functions to numerically integrate the evolution equations (originally done by \cite{Page:1976ki,Chambers:1997ai,Taylor:1998dk}; we use those computed in \cite{Calza:2023rjt}). For example, the full $e_T$ including all Standard Model and graviton degrees of freedom is approximately given by (for $a\lesssim 0.6$):
	\begin{align}
		e_T \simeq {0.0044 \over 1- a^2} \left( 1+ 0.5 \, a \right) \label{eTSM}
	\end{align}
	and we plot $e_T$ together with the spin-0 contribution $f_0$ in Fig.~\ref{fig:eTandf0}. Note that a single dark matter boson or an $\mathcal{O}(1)$ number of string axions do not contribute much to $e_T$, but with many axions, $N_a\gtrsim 100$, axion emission may become significant (or even dominant) depending on the initial spin parameter. As we will see later, this leads $N_a \sim \mathcal{O}(100)$ to be particularly interesting to black hole evaporation and superradiant cloud formation.
	
	Unlike the emission of particles with non-zero spin, scalar emission may be spherically symmetric (in the monopole, $\ell=0$ mode), therefore decreasing a PBH's mass but not its angular momentum and, hence, increasing the spin parameter $a$. Previous work \cite{Calza:2021czr} included $N_a$ species of axion-like particles in the Hawking emission spectrum and found that the black hole evaporation dynamics can be significantly modified, with PBHs surviving until the present day (born with masses around $10^{11}-10^{12}$ kg) developing significant spin parameters (up to $a\simeq 0.555$) even if born with negligible spin. In fact, finding PBH remnants with masses $\sim 10^{10}-10^{11}$ kg would be evidence for hundreds of light axion species. That work also predicted a significant present abundance of relativistic axions from PBH evaporation. In this paper, we consider instead a population of PBHs that evaporate completely before nucleosynthesis, with no black holes surviving until today. In this way, potential signatures of a large number of axions in the scenario we study here complement these earlier results.

	\subsection{Evaporation with superradiance}
	
	For bosonic particles with Compton wavelength comparable to the black hole horizon, scattering off the horizon allows the field to extract energy and angular momentum from the black hole. If the particles are gravitationally bound to the black hole, this results in a ``superradiant instability'' where the particle number in the gravitationally bound cloud grows exponentially fast. For a boson with mass $\mu$, this can only occur when the gravitational coupling is small,
	\begin{align}
		\alpha \equiv {\mu M \over \mpl^2} \lesssim 1 ~.
	\end{align}
	
	Bosonic quasi-bound states around a black hole are described by quantum numbers $(n,j,\ell,m)$, where the principal quantum number $n=\ell+1+n_r$, with $n_r$ denoting the radial node number. Each state is characterized by orbital angular momentum numbers $(\ell,m)$, with $m$ corresponding to the projection along on the black hole's rotation axis, and by a total angular momentum $j$. For scalar fields $\ell=j$, while for vector fields $\ell=j-1,j,j+1$. In both cases quasi-bound states have complex frequencies, where the real part exhibits a Hydrogen-like spectrum:
	\begin{align}
		\omega_\text{R} \simeq \mu \left(1- {\alpha^2 \over 2n^2}\right)~.
	\end{align}
	The superradiant instability occurs for $\omega_\text{R} < m \Omega$, where $\Omega= a \mpl^2/[2M(a+\sqrt{1-a^2})]$, for which the imaginary part of the frequency of a given state becomes positive, with $\omega_\text{I}\propto m\Omega-\omega$. Since this leads to an exponential growth of the particle number at a rate $\Gamma_\text{S}=2\omega_\text{I}$, the resulting boson cloud is completely dominated by the fastest growing mode, with the number of particles in the cloud found by integrating
	\begin{align}
		{d N_\text{S} \over dt} = \Gamma_\text{S}\left(M,a,\mu\right) N_\text{S}~, \label{dNdt}
	\end{align}
	for a single mode. In the non-relativistic regime, $\alpha <1$, the superradiant growth rate for the dominant scalar mode ($\ell=m=1$) is approximately given by: 
	\begin{align}
		\Gamma_\text{S} \simeq {1\over 24} \left(a - 4\alpha\right)\alpha^8 \mu~. \label{GammaApprox}
	\end{align}
	In the simulations described in the next section, we use the more precise expression derived from numerical results in \cite{Calza:2023rjt}. The superradiant condition can be read off as $\alpha<a/4$ (equivalent to $\omega_\text{R} < m \Omega$ above). Very light bosons, such as the string axions we will be interested in, satisfy the superradiant condition. However, the growth rate Eq.~(\ref{GammaApprox}) comes with a factor of $\alpha^8$, meaning the superradiant cloud will essentially be only dark matter particles (with $\alpha \lesssim 1$) and no axions ($\alpha \ll1$).
	
	Each particle produced by the superradiant instability extracts its mass and spin from the black hole, so we must update Eq.~(\ref{evapeqs}) to
	\begin{align}
		{d M \over d t} &= - e_T {\mpl^4 \over M^2} - \mu_{} \Gamma_\text{S} N_\text{S}   \label{dMdt} \\
		{d J \over d t} &= -e_J { J \mpl^4 \over M^3} - \Gamma_\text{S} N_\text{S}  \label{dJdt}
	\end{align}
	and the evolution of the system of black hole and superradiant cloud is determined by Eqs.~(\ref{dNdt}), (\ref{dMdt}), and (\ref{dJdt}). As the black hole evaporates and spins down, $\Gamma_\text{S}$ can change sign; When $\Gamma_\text{S}<0$, the bound state decays away from the superradiant regime, signaling their absorption by the black hole.    
	In fact, $N_\text{S}$ corresponds to the expectation value of a quantum number operator, $\langle N_\text{S}\rangle$, and quantum fluctuations prevent this from becoming arbitrarily small \cite{Kofman:1982gu,Fu:2025ztk}, even if classically it decays exponentially outside the superradiant regime. For this reason we impose $N_\text{S}\geq1$ in our (classical) simulations, detailed in the following section.
	
	The dynamical interplay between Hawking emission and superradiant dark matter production is best understood by considering the dynamical equation for the dimensionless spin parameter, which can be derived from Eqs.~(\ref{dMdt}) and (\ref{dJdt}):
	\begin{equation}
		{da\over dt}=a(2e_T-e_J){\mpl^4\over M^3}-\Gamma_\text{S}N_\text{S}(1-2\alpha a){\mpl^2\over M^2}~.   
	\end{equation}
	In the Standard Model, $e_J> 2e_T$ and evaporation makes a black hole spin down as it evaporates. Hence, even if dark matter bosons initially satisfy the superradiance condition, a dark matter cloud can only grow efficiently if superradiance is much faster than evaporation. In this case $N_\text{S}$ grows, depleting the black hole spin until the superradiance condition is saturated, $\omega\simeq \Omega$ (for $\ell=m=1$), reaching a maximum value $N_\text{S}^\text{max}\sim a (M/\mpl)^2$ after $\mathcal{O}(100)$ e-folds. An efficient superradiant dark matter production therefore requires $\Gamma_\text{S}t_\text{evap}\gtrsim 100$ \cite{March-Russell:2022zll}, where $t_\text{evap}\simeq M_0^3/(3e_T \mpl^4)$ is the black hole's evaporation time for an initial mass $M_0$. 
	
	However, even if the cloud attains its maximum mass, Hawking emission continues to deplete the black hole spin, eventually making $\Gamma_\text{S}<0$ and triggering the decay (i.e.~reabsorption) of the cloud. Given the dependence of $\Gamma_\text{S}$ on the mass coupling $\alpha$, which decreases as the black hole evaporates, this reabsorption becomes more and more suppressed, yielding a constant number of particles in the cloud. The analysis in \cite{March-Russell:2022zll} identified an optimal regime for which superradiant dark matter production is efficient, with little reabsorption, for PBHs with percent-level spin parameters in the mass range $M_0\sim 10^5-10^6$ kg and dark scalar bosons with $\mu\sim 1-10$ TeV, corresponding to $\alpha\sim \mathcal{O}(10^{-3})$ (and much broader parametric ranges for larger black hole spins and for spin-1 dark bosons). We will show in the next section that this changes with the inclusion of light axions in the Hawking emission spectrum, which can slow the black hole's spin-down or even reverse it, depending on the number of species.

	\begin{figure}[h]
		\centering
		\includegraphics[width=\columnwidth]{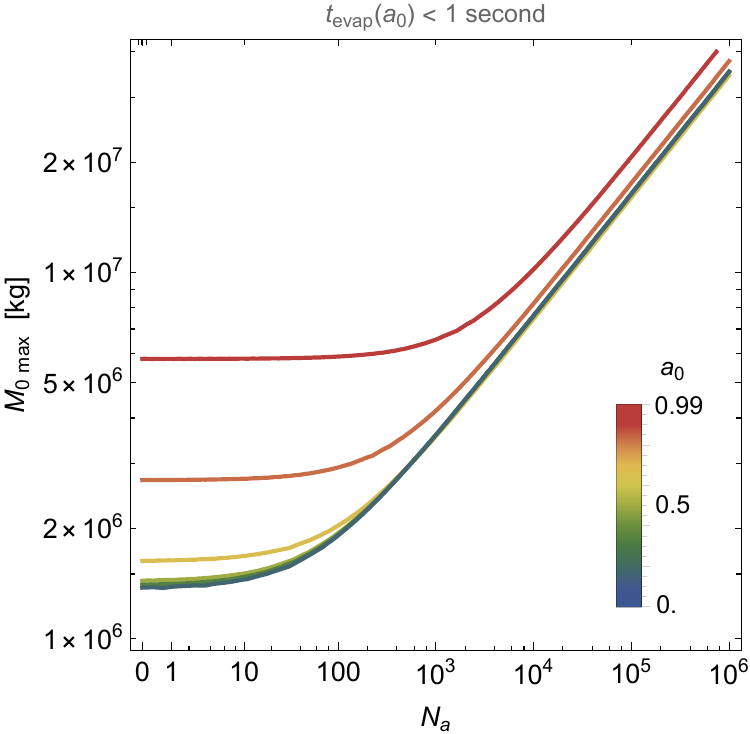}
		\caption{Maximum initial mass of PBHs that evaporate before BBN, $t_\text{\tiny{BBN}} \sim 1$ second, for different numbers of axion species.}
		\label{fig:tevap}
	\end{figure}
	
	Note that in this scenario dark matter is produced by both Hawking emission and superradiance, seen by comparing the dark matter mass to the PBH temperature,
	\begin{align}
		{T_\text{H} \over \mu} = {1 \over 4 \pi \alpha} { \sqrt{1-a^2} \over \left( 1 + \sqrt{1-a^2} \right)} ~. 
	\end{align}
	For small PBH spins, $T_\text{H}\gtrsim \mu$ for $\alpha\lesssim 0.04$, so Hawking emission occurs within the parametric regime relevant for superradiance. Reference \cite{March-Russell:2022zll} showed that, for percent-level PBH spins, superradiance may account for up to $\mathcal{O}(10\%)$ of the number of dark matter particles produced by each black hole in the scalar case, becoming the dominant production mechanism at larger values of the PBH spin parameter. We will also see in the next section that a much larger fraction of dark matter can be produced by superradiance in the context of the string axiverse.

	\section{Including any number of axions}\label{sec:Na}

	In this section we include an arbitrary number of axion species, $N_a$, in the PBH Hawking emission spectrum to study its effect on the superradiant production of dark matter bosons. We assume that all axions are very light, and may be treated as massless, in contrast to the heavy dark matter scalar. Note that, in this case, superradiant axion production is very suppressed, recalling that $\Gamma_\text{S}\propto \alpha^8$. Hence, only the heavy dark matter boson is produced by PBH superradiance, whereas both the latter and the light axions are produced by Hawking emission.
	
	Black hole evaporation is controlled by the total mass-emissivity coefficient, given by
	\begin{align}
		e_T(a) = \bar{e}_T(a)+N_a \, f_0(a)
	\end{align}
	where $\bar{e}_T(a)$ includes the contribution of SM particles, the graviton and the dark matter boson. Similarly, the spin-emissivity coefficient is given by
	\begin{align}
		e_J(a) = \bar{e}_J(a)+ N_a \, g_0(a) ~.
	\end{align}   
	Including multiple new spin-0 emission channels speeds up PBH evaporation, with the evaporation time approximately given by
	\begin{align}
		t_\text{evap} \simeq {M_0^3 \over 3 e_T(a_0) \mpl^4} \label{tevap}
	\end{align}
	in terms of the initial black hole mass $M_0$, and can exceed $t_\text{\tiny{BBN}}$ for sufficiently heavy $M_0$. Fig.~\ref{fig:tevap} shows the maximum PBH formation mass for which the black hole evaporates within 1 second, as a function of $N_a$.
	
	Standard PBH formation scenarios, where overdense regions re-enter the Hubble horizon and collapse in the radiation-dominated era, are expected to yield PBH spin parameters at or below the percent level, so we are primarily interested in the lowest curves of Fig.~\ref{fig:tevap}. The evaporation time is not strongly affected by $N_a$, as $M_{0\,\text{max}}$ only changes by an order of magnitude over six decades of $N_a$. As we will see below, superradiant dark matter production is more efficient for heavier black holes, and in the following plots we frequently take $M_0 = 5 \times 10^5$ kg as a reference value, near but safely below the upper limit set by $t_\text{evap} < t_\text{\tiny{BBN}}$ for all $N_a$ and $a_0$.

	\subsection{Dynamics of superradiance and evaporation}\label{sec:dynamics}
	
	%%% dynamical plots
	\begin{figure}
		\centering
		\includegraphics[width=\columnwidth]{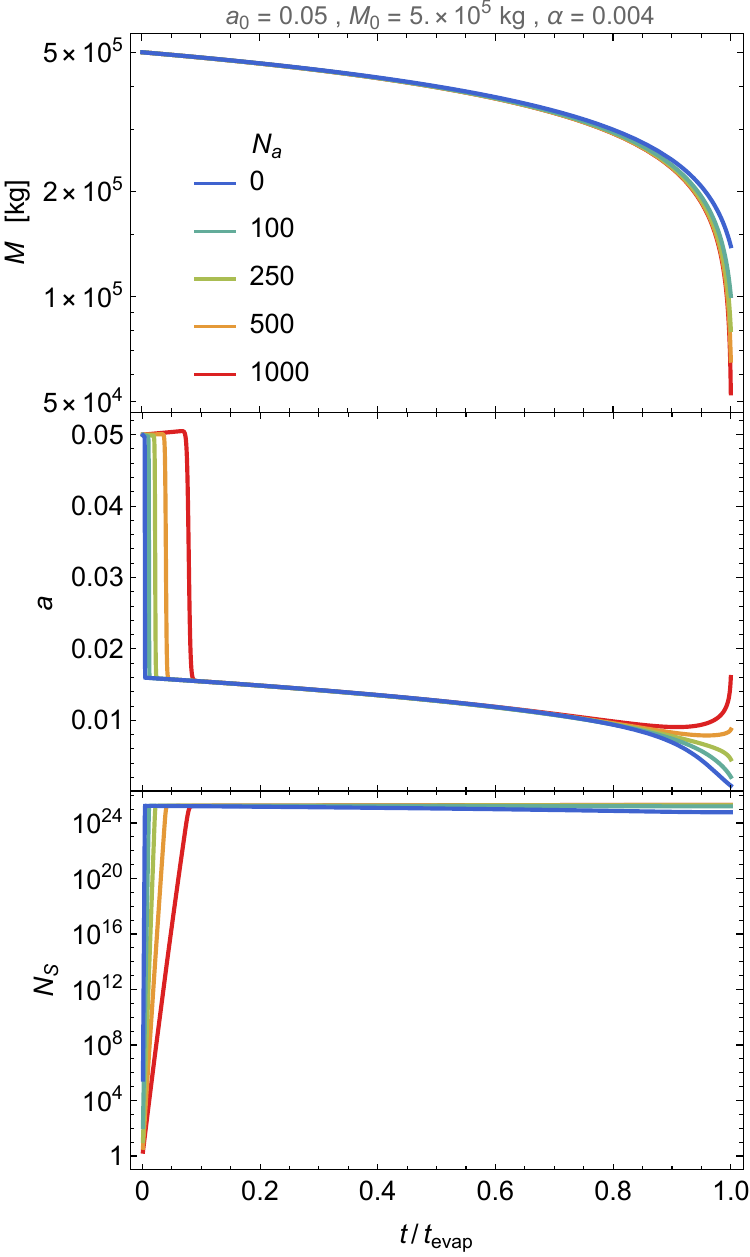}
		\caption{Black hole mass and spin, and number of particles in the superradiant cloud, for choices of $a_0 = 0.05$ and $\alpha = 0.004$. Note that $t_\text{evap}$, approximated by Eq.~(\ref{tevap}), is different for each curve.}
		\label{fig:dyna05}
	\end{figure}
	\begin{figure}
		\centering
		\includegraphics[width=\columnwidth]{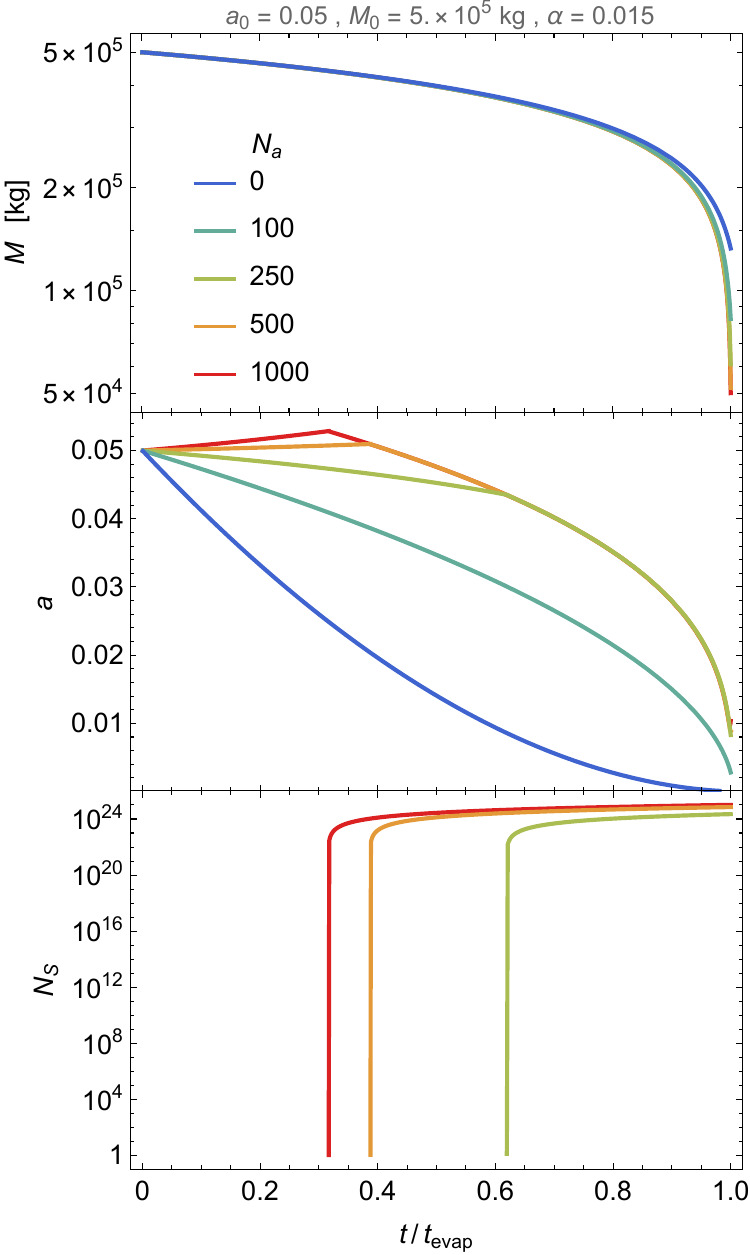}
		\caption{Same initial spin as Fig.~\ref{fig:dyna05}, $a_0 = 0.05$, and larger $\alpha = 0.015$.}
		\label{fig:dyna05la}
	\end{figure}
	\begin{figure}
		\centering
		\includegraphics[width=\columnwidth]{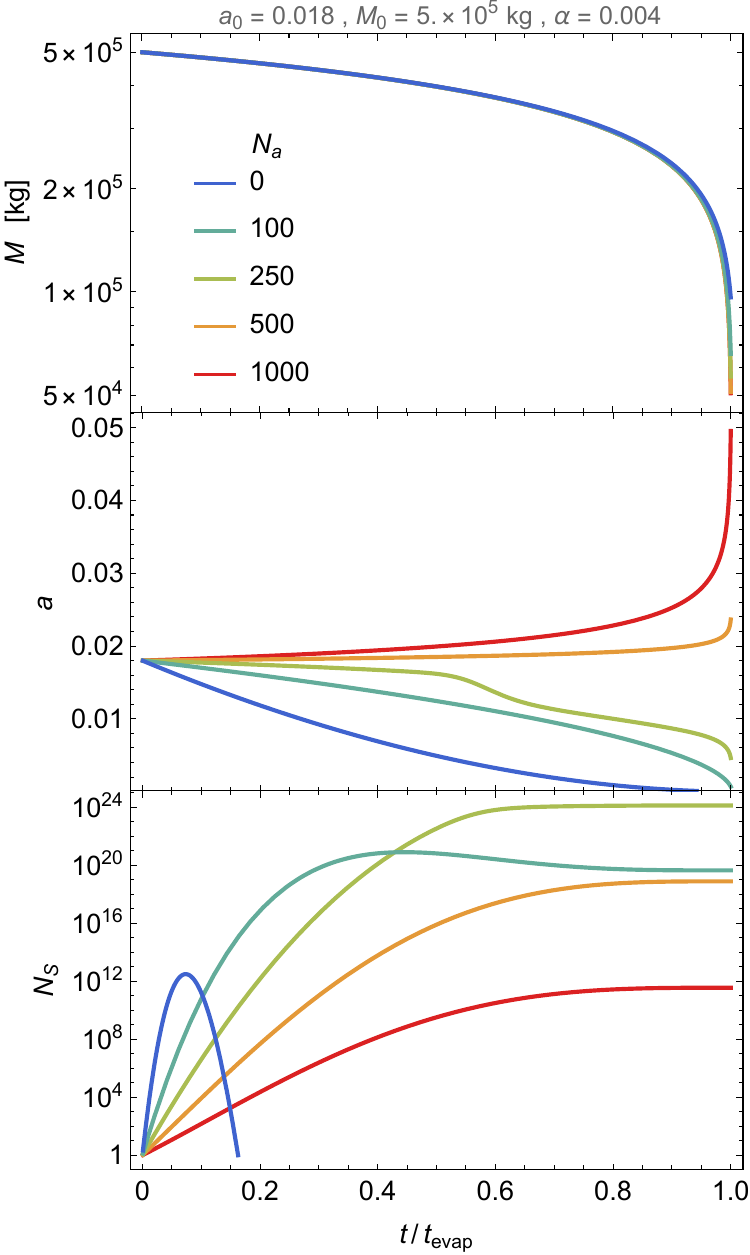}
		\caption{Lower initial spin, $a_0 = 0.018$, and smaller $\alpha = 0.004$.}
		\label{fig:dyna018}
	\end{figure}
	\begin{figure}
		\centering
		\includegraphics[width=\columnwidth]{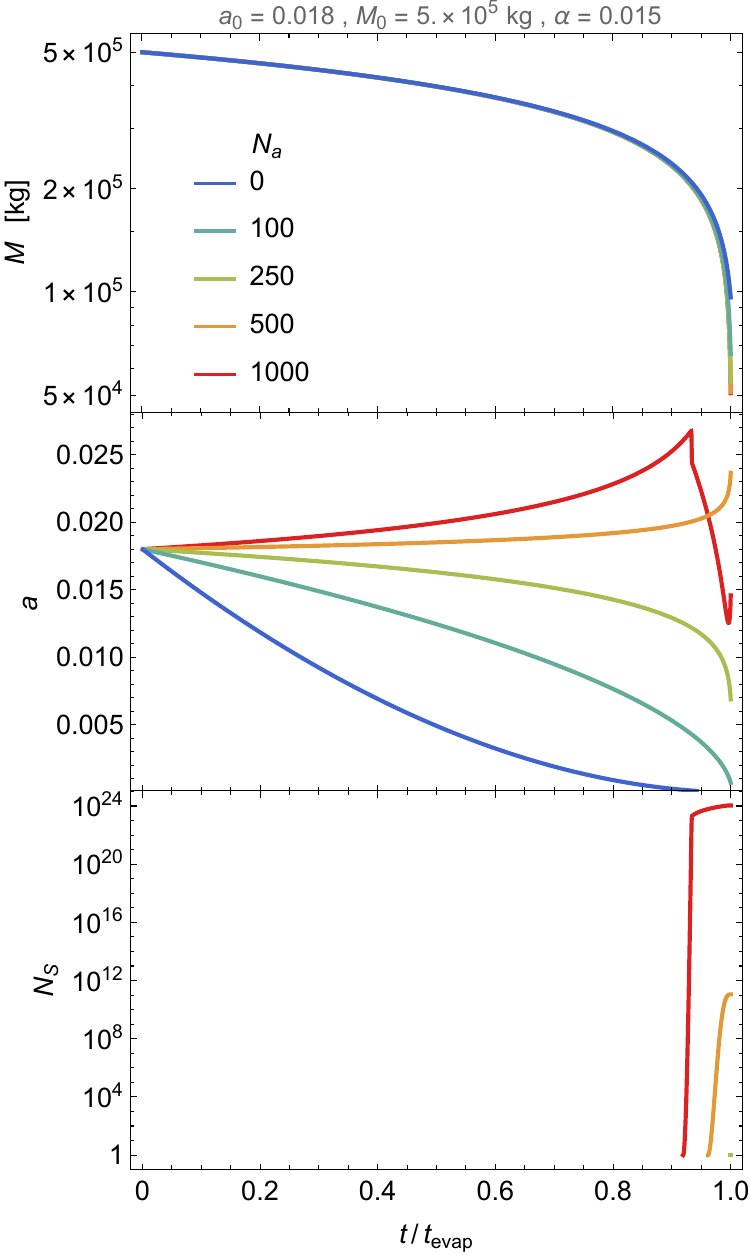}
		\caption{Lower initial spin, $a_0 = 0.018$, and larger $\alpha = 0.015$.}
		\label{fig:dyna018la}
	\end{figure}

	We have numerically evolved the PBH and superradiant dark matter cloud system, according to Eqs.~(\ref{dNdt}), (\ref{dMdt}), and (\ref{dJdt}), for several choices of the PBH's initial spin $a_0$, and mass $M_0$, focusing on $a_0 \leq 0.05$ for the above-mentioned reasons. A selection of results is given in Figs.~\ref{fig:dyna05}--\ref{fig:dyna018la}, where we show the evolution of the PBH mass and dimensionless spin, and the number of dark matter particles in the superradiant cloud, for $M_0=5\times 10^5$ kg. 
	
	While a large number of axion species may spin up the PBH, this should not be conflated with a growth in $N_\text{S}$, and spin-up is not a requisite for superradiant cloud formation. In particular, the onset of superradiance always spins down the PBH, even for large $N_a$; and even in cases where the PBH spins up, the superradiant cloud does not always attain its maximum occupation number.
	
	In these figures, we consider two initial PBH spins, $a_0 = \lbrace0.05,~ 0.018\rbrace$, and two values of $\alpha\equiv\mu M_0 /\mpl^2 = \lbrace 0.004,~ 0.015\rbrace$ (or equivalently, dark matter mass $\mu = \lbrace2.1~\text{TeV},~7.9~\text{TeV}\rbrace)$. The initial number of particles in the superradiant cloud is set to $N_\text{S}=1$.
	
	There are four qualitatively different evaporation scenarios: (1) In the simplest case, the superradiance condition is not initially satisfied, and $N_\text{S}$ decays immediately: a dark matter cloud is not produced at all (curves with smaller values of $N_a$ in Figs.~\ref{fig:dyna05la} and \ref{fig:dyna018la}). (2) If the superradiance condition is initially satisfied, $N_\text{S}$ will immediately grow exponentially. For appropriate parameters, even in the $N_a = 0$ case this can lead to a significant dark matter cloud after the black hole has evaporated, with $N_\text{S}\rightarrow N_\text{S}^\text{max}\sim 10^{25}$, as in Fig.~\ref{fig:dyna05}. 
	
	(3) For heavier dark matter particles (i.e.~larger $\alpha$), there can be a significant reabsorption of the superradiant cloud by the PBH. In Fig.~\ref{fig:dyna05} a mild reabsorption can be seen in the $N_\text{S}$ plot, where the $N_a = 0$ curve ends somewhat below those corresponding to greater numbers of axion species. This behavior is more pronounced in Fig.~\ref{fig:dyna018}, for a lower initial PBH spin. There, in the $N_a =0$ case the superradiant cloud is completely reabsorbed before $t = 0.2 \, t_\text{evap}$. With $N_a=100$, for which the PBH evaporates slightly faster, less reabsorption of the superradiant cloud occurs within $t_\text{evap}$. Sufficiently many axion species eliminates reabsorption entirely, but even greater $N_a$ provides less time for the superradiant cloud to build up before the black hole completely evaporates, resulting in lower final occupation number $N_\text{S}$.
	
	(4) In the examples shown in Figs.~\ref{fig:dyna05la} and \ref{fig:dyna018la} with larger values of $N_a$, the superradiant cloud grows and reaches its maximum mass even though the superradiance condition is not initially satisfied. In these cases the PBH spin down is delayed or even inverted, such that $a>4\alpha$ before the PBH evaporates significantly, thus triggering the superradiant instability. 
	Since the classical evolution of Eqs.~(\ref{dNdt}), (\ref{dMdt}), and (\ref{dJdt}) leads to $N_\text{S}\ll1$ outside the superradiant regime, we add 1 to $N_\text{S}$ if $a$ ever grows larger than $4\alpha$ to accurately model its evolution in these cases, taking into account quantum effects discussed earlier.
	In the cases where the Hawking emission spins up the PBH, the superradiance condition is never violated and eventually saturates, so reabsportion never occurs. This dynamical scenario is only possible for $N_a \gtrsim \mathcal{O}(100 - 1000)$ axions (depending on $a_0$, $M_0$, and $\alpha$), thus showing that the string axiverse considerably broadens the parameter space for efficient superradiant dark matter production, as we will see in more detail below.

	\subsection{Efficiency of superradiant DM production}\label{sec:efficiency}

	Let us define the efficiency of superradiant dark matter production as the ratio of superradiant particle production to the total dark matter production including Hawking emission:
	\begin{align}
		\epsilon \equiv  \left({ N_\text{S} \over N_\text{S}+N_\text{HE}}\right)\bigg{|}_{t_\text{evap}} ~. \label{efficiency}
	\end{align}
	Evaluating at $t_\text{evap}$ determines the total dark matter production, as the superradiant cloud can be (partially) reabsorbed during evaporation. Previous work studying the $N_a =0$ case found a maximum efficiency $\epsilon_\text{max} = 13\%$ ($73\%$) for spin-0 (spin-1) dark matter at initial black hole spin $a_0=0.05$, with significantly greater efficiency for larger $a_0$ (possible in e.g.~PBH formation in an early matter-dominated epoch). 
	
\begin{figure}
		\centering
		\includegraphics[width=\columnwidth]{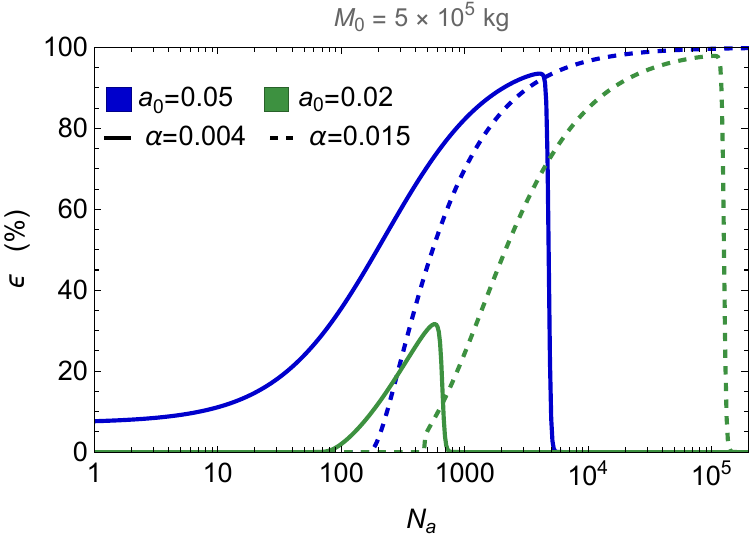}
		\caption{Efficiency $\epsilon$ as a function of the number of axion species $N_a$, for several choices of initial PBH spin $a_0$ and gravitational coupling $\alpha$, with initial mass $M_0=5\times 10^5$ kg near the limit illustrated in Fig.~\ref{fig:tevap}. For $a_0=0.05$ (upper solid curve),  superradiance can be efficient without axions but is much more efficient with greater $N_a$. With $a_0=0.02$ (lower solid curve), $\epsilon>1\%$ only for a range of $N_a \sim 100 - 700$. Higher $\alpha$ (dashed curves) have somewhat greater $\epsilon$, over a larger range of $N_a$.}
		\label{fig:effcurves}
	\end{figure}

	We have numerically computed this efficiency for different values of the PBH mass and spin and the dark matter mass, and in Fig.~\ref{fig:effcurves} we show a few examples as a function of $N_a$, with $M_0 = 5\times 10^5$ kg as before. Larger black hole masses also roughly correspond to higher efficiency, so this choice of $M_0$ is illustrative in showing the improvement to the efficiency as a function of $N_a$.
	
	For $a_0=0.05$ (blue curves in Fig.~\ref{fig:effcurves}), superradiance can be efficient with $N_a=0$ for the choice of $\alpha=0.004$ (i.e.~$\mu=2.1$ TeV), but is much more efficient with $N_a \sim \mathcal{O}(100 - 10^3)$. For mildly heavier dark matter, $\alpha=0.015$ (i.e.~$\mu=7.9$ TeV) $\epsilon$ can even approach 100$\%$ for sufficiently many axion species, $N_a \sim \mathcal{O}(10^4 - 10^5)$. For smaller initial black hole spin, $a_0=0.02$ (green curves in Fig.~\ref{fig:effcurves}), superradiance produces essentially none of the dark matter if $N_a=0$. However, the situation is greatly improved  with a large number of axion species, for which even the $a_0=0.02$ cases can reach significant efficiencies. 

	\begin{figure}
		\centering
		\includegraphics[width=\columnwidth]{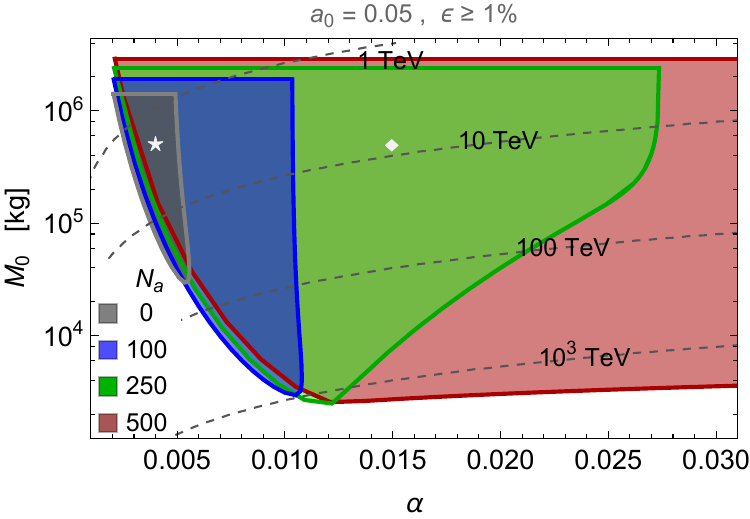}
		\caption{Regions of mass parameter space with at least 1\% efficiency of superradiant DM production, for several choices of number of axion species $N_a$, with initial black hole spin $a_0=0.05$. Values of the DM particle mass $\mu = \alpha \mpl^2 / M_0$ are shown as dashed contours; star and diamond markers indicate the masses used in Figs.~\ref{fig:dyna05} -- \ref{fig:effcurves}. The upper limit of $M_0$ for each $N_a$ is set by requiring $t_\text{evap} < 1$ second $\sim t_\text{\tiny{BBN}}$. Increasing $N_a$ greatly expands the efficient parameter space to larger $\alpha$ and lower $M_0$.
		}
		\label{fig:1percent}
	\end{figure}
	
	While lower $a_0$ generally leads to smaller $\epsilon$ (in Fig.~\ref{fig:effcurves} the green curves lie below the corresponding blue curves), nevertheless the smaller spin case can also have significant efficiency in a range of $N_a$. The range of $N_a$ with non-zero efficiency increases with $\alpha$, meaning that there is wider parametric range to produce heavier dark matter through superradiance, compared to the case $N_a=0$.
	
	In each $\epsilon(N_a)$ curve shown in Fig.~\ref{fig:effcurves}, there is a maximum $N_a$ for which $\epsilon \to 0$ (note that the sharpness of this drop-off is simply due to the logarithmic plot scaling). Above this maximum value, the PBH evaporates (mostly into axions) faster than superradiance builds a dark matter cloud. This behavior can also be seen in Fig.~\ref{fig:dyna018}, where increasing $N_a>250$ leads to fewer particles in the superradiant cloud. Note that this maximum number of axion species depends on $a_0$, $M_0$, and $\alpha$.

	Figure \ref{fig:1percent} shows the range of PBH and dark matter masses $\lbrace M_0,\,\alpha \rbrace$ with at least 1$\%$ efficiency, for several values of $N_a$. While this shows only the $\epsilon=1\%$ contour, without showing the gradient inside the shaded regions, we note that the efficiency can reach values as large as $\epsilon \gtrsim 50\%$ for $N_a = 500$. The $\epsilon$ gradient within these regions is similar to the $N_a = 0$ case studied in \cite{March-Russell:2022zll}, with the maximum efficiency found at small $\alpha$ and large $M_0$ inside each shaded region (near the star marker, upper left of Fig.~\ref{fig:1percent}). 
	
	The largest number of axions we show here is $N_a = 500$; further increasing $N_a$ expands the $\epsilon \geq 1\%$ region to even higher $\alpha$ (where in principle one should take into account $\mu>T_\text{H}$ when computing $\epsilon$). Of primary interest in Fig.~\ref{fig:1percent} is the effect of increasing $N_a$: an enormous expansion in the mass parameter space of efficient superradiant dark matter production compared to the $N_a =0$ case. Alongside the increase in efficiencies possible with small PBH spins, illustrated in Fig.~\ref{fig:effcurves}, these are our main results.

	\section{Axion contribution to $\Delta N_\text{eff}$}\label{sec:DeltaNeff}
	
	As discussed earlier, the hot axions emitted by the PBH behave as dark radiation\footnote{Here we are assuming the typically large axion decay constants of string axions, not far from the GUT scale, which prevents them from reaching thermal equilibrium with the SM plasma.}. Having considered in previous sections as many as $N_a\gtrsim 10^5$ species of axions, one may be concerned that so many additional light degrees of freedom will violate observational bounds from the CMB on the effective number of relativistic degrees of freedom, usually put in terms of an effective number of neutrinos, $N_\text{eff}$ \cite{Planck:2018vyg} \footnote{Hawking gravitons also contribute to $N_\text{eff}$ \cite{Hooper:2019gtx,Hooper:2020evu,Arbey:2021ysg,Franciolini:2026fdv,Jia:2026psi}, although this is suppressed for low PBH spins and/or in scenarios where the PBHs never dominate the energy density before evaporating away, as in the present case (see Sec.~\ref{sec:RhoPBH}). } 
	
	This will not, however, be a concern for the scenario studied in this paper. The PBH number density is fixed by the amount of dark matter that each produces (via superradiance and Hawking emission) and the presently inferred dark matter density. This in turn fixes the axion production by Hawking emission in terms of the PBH natal mass $M_0$ and spin $a_0$, the dark matter mass $\mu$ and the number of axion species $N_a$. The maximum $\Delta N_\text{eff}$ we find at an intermediate $N_a$ is quite small for well-motivated choices of the mass parameters. Further increasing $N_a$, $\Delta N_\text{eff} \sim N_a^{-1/2}$ (one may see ahead to Eq.~\ref{DNeffApprox}). That is, rather than growing without bound, $\Delta N_\text{eff} \to 0$ for sufficiently large $N_a$.

	\subsection{Axion energy density compared to SM radiation}
	
	To track the relevant number densities without following their cosmological evolution, it is convenient to consider the corresponding yields $Y_i \equiv {n_i / s}$, where $n_i$ is the number density of component $i$, and $s$ is the SM entropy density, so that $Y_i$ is independent of the scale factor. The number of axions emitted during evaporation is set by 
	\begin{align}
		Y_a = Y_\text{BH} \, N_{0} \, N_a
	\end{align}
	where $N_{0}$ is the number of Hawking-emitted relativistic massless axions from each PBH, and $N_a$ is the number of axion species. Similarly, the dark matter yield is
	\begin{align}
		Y_\text{DM} = Y_\text{BH} \, N_\text{DM} \label{YDM}
	\end{align}
	where the number of dark matter particles produced by each black hole comes from both Hawking emission and superradiance (evaluated at $t_\text{evap}$),
	\begin{align}
		N_\text{DM} &=  N_\text{HE} + N_\text{S} \equiv {N_\text{S} \over \epsilon} \equiv {N_\text{HE} \over 1-\epsilon}
	\end{align} using the superradiant efficiency defined in Eq.~(\ref{efficiency}).
	Thus,
	\begin{align}
		Y_a &= Y_\text{DM} \, {N_{0} \over N_\text{DM}} \, N_a \\
		&= \left(1-\epsilon \right) Y_\text{DM} \, N_a \label{Ya}
	\end{align}
	where in the second line we specialize to the case where the dark matter is also spin-0, $N_{\text{HE}}=N_{0}$ (for $\alpha\lesssim 0.04$ such that $T_H\gtrsim \mu$, as in the previous section), thereby fixing the total axion yield in terms of the dark matter yield, which is fixed by the present dark matter abundance to be $\mu \, Y_\text{DM} = 0.43$ eV (i.e.~$\Omega_\text{DM} h^2=0.11$). 
	
\begin{figure}
		\centering
		\includegraphics[width=\columnwidth]{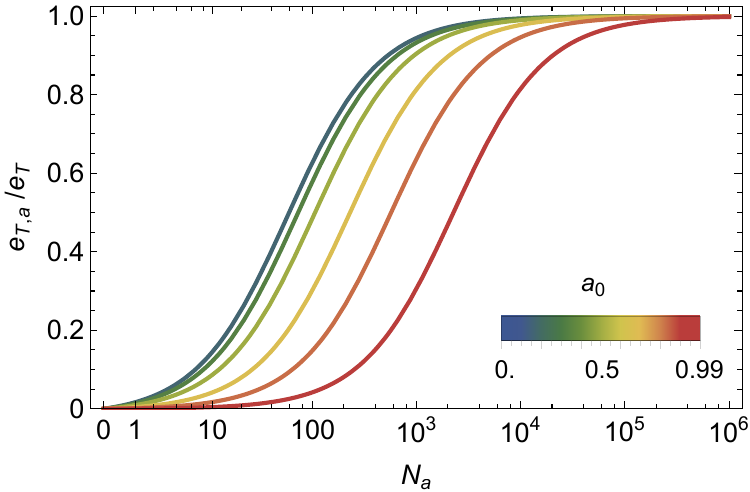}
		\caption{The ratio of axion to total emissivity coefficients is independent of $a_0$ for low initial spins, $a_0 \lesssim 0.1$. In the large $N_a$ limit, evaporation is dominated by axion emission.}
		\label{fig:eTratio}
	\end{figure}

	The axion number density at the time of PBH evaporation is determined by the SM entropy density, $n_a(t_\text{evap}) = Y_a \times  s(T_\text{evap})$, where the entropy at $T_\text{evap}$ during radiation domination is
	\begin{align}
		s\left(T_\text{evap}\right) &= {2\pi^2 \over 45} g_\star\!\left(T_\text{evap}\right) T_\text{evap}^3 \\
		&= {2\pi^2 \over 45} g_\star\!\left(T_\text{evap}\right) \left( {45 \over 16 \pi^3 g_\star} {\mpl^2 \over t_\text{evap}^2} \right)^{3/4} \label{sevap}
	\end{align}
	with the evaporation time given in Eq.~(\ref{tevap}) fixed to be earlier than BBN. The energy per axion is the fraction of a PBH's initial mass emitted into axions,
	\begin{align}
		\mathcal{E}_a(t_\text{evap}) = {M_0 \over N_0 N_a} {e_{T,a} \over e_T} \label{Eaxion}
	\end{align}
	where $e_{T,a}=f_0 N_a$ is the emissivity coefficient of only the axions (with $N_a$ species), and $e_T$ is the total emissivity coefficient. This ratio of emissivities, plotted in Fig.~\ref{fig:eTratio}, varies from 0 to 1 with $N_a$ and is effectively independent of the black hole initial spin at the low spins in which we are mainly interested, $a_0 \leq 0.05$. It is interesting to note that for low spins, with $N_a \sim \mathcal{O}(100)$, axions already comprise $60\% - 90\%$ of Hawking emission. 
	
	Combining Eqs.~(\ref{Ya}), (\ref{sevap}), and (\ref{Eaxion}), the energy density of axions at the time of black hole evaporation is then
	\begin{align}
		\rho_a(t_\text{evap}) &= \mathcal{E}_a(t_\text{evap}) \,n_a(t_\text{evap}) \\
		&= {M_0 \over N_0 N_a} {e_{T,a} \over e_T} \left(1-\epsilon \right) Y_\text{DM} \, N_a \nonumber \\
		&~~~\times {2\pi^2 \over 45} g_\star \left( {45 \over 16 \pi^3 g_\star} {\mpl^2 \over t_\text{evap}^2} \right)^{3/4} ~.
	\end{align}
	The SM radiation energy density at $T_\text{evap}$ is 
	\begin{align}
		\rho_\text{R}\left(T_\text{evap}\right) = {\pi^2 \over 30} g_\star T_\text{evap}^4 = {3 \over 32 \pi} \left({\mpl \over t_\text{evap}}\right)^2 ~.
	\end{align}
	At matter-radiation equality, the ratio of axion-SM radiation energy densities evolves to
	\begin{align}
		{\rho_a \over \rho_\text{R}}\bigg{|}_\text{eq} = {\rho_a \over \rho_\text{R}}\bigg{|}_\text{evap} \left({g_{\star \text{evap}} \over g_{\star \text{eq}} }\right) \left({ g_{\star S\,\text{eq}} \over g_{\star S\,\text{evap}} }\right)^{4/3}
	\end{align}
	and the resulting change in the effective number of neutrino species fit to CMB observations is
	\begin{align}
		\Delta N_\text{eff} = {\rho_a \over \rho_\text{R}}\bigg{|}_\text{eq} \left(\frac{8}{7} \left(\frac{11}{4}\right)^{4/3}+N_\nu\right)
	\end{align}
	where we take the observed effective number of neutrino generations to be $N_\nu =3.046$. At $T_\text{evap} \sim \mathcal{O}(10-100)$ MeV ($t_\text{evap} < 1$ second), the photons, neutrinos, and electrons in equilibrium give $g_\star (T_\text{evap})= g_{\star S}(T_\text{evap}) = 10.75$. At matter-radiation equality, with $T_\text{eq} \lesssim 1$ eV, we have $g_\star\left(T_\text{eq}\right) \simeq 3.36$ and $g_{\star S}\left(T_\text{eq}\right) \simeq 3.91$, which gives $	({\rho_a / \rho_\text{R}})|_\text{eq} \simeq 0.77 \times ({\rho_a / \rho_\text{R}})|_\text{evap} $. 
	
\begin{figure}
		\centering
		\includegraphics[width=\columnwidth]{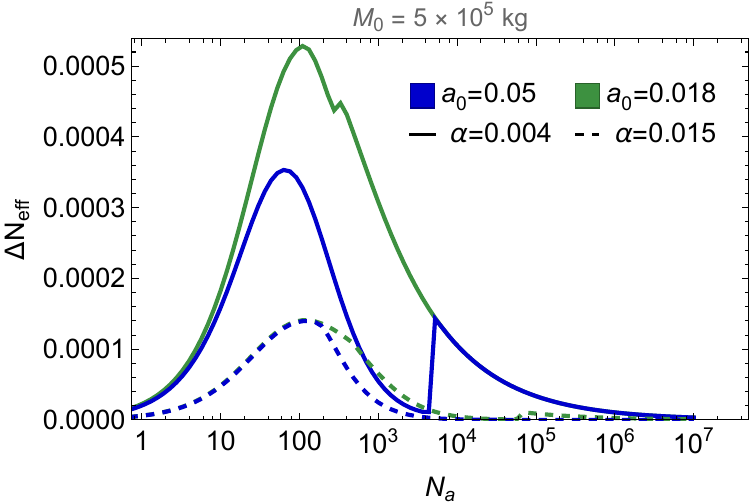}
		\caption{Change to the effective number of neutrino species $\Delta N_\text{eff}$ due to $N_a$ species of axions, for several choices of (low) initial black hole spin $a_0$ and coupling $\alpha\equiv M_0 \mu /\mpl^2$. Rather than monotonically increasing with the number of axions, $\Delta N_\text{eff}$ has a maximum around $N_a \sim \mathcal{O}(100)$, which is always below the observational bound of $\Delta N_\text{eff} \lesssim 0.1$.}
		\label{fig:DeltaNeff}
	\end{figure}

	Altogether we obtain approximately
	\begin{align}
		\Delta N_\text{eff} &\simeq 25.23 {e_{T,a} \over e_T} {M_0 \over N_0} \sqrt{t_\text{evap} \over \mpl} Y_\text{DM} \left(1-\epsilon\right) \label{DeltaNeff} \\
		&\simeq {0.0193 N_a \over (59.8+N_a)\sqrt{44.6+0.744 N_a} } \left(1-\epsilon\right) \nonumber \\
		&~~~\times \left( {M_0 \over 5\times 10^5 \text{ kg}} \right) \left( {1 \text{ TeV}\over\mu } \right) \label{DNeffApprox}
	\end{align}
	where the second line is valid for $a_0<0.1$, having approximated the emissivities, $N_0$, and evaporation time at low spin for which they are independent of $a_0$. Very importantly, the efficiency $\epsilon\in [0,1]$ depends on $M_0$, $a_0$, and $N_a$ as detailed in Sec.~\ref{sec:efficiency}, though we do not have a semi-analytical expression for $\epsilon$. We include the full black hole spin dependence to numerically compute and plot $\Delta N_\text{eff}$ as a function of $N_a$ in Fig.~\ref{fig:DeltaNeff}. The maximum value curiously occurs with roughly hundreds of axion species, $N_a \sim \mathcal{O}(100)$, the number of axions expected in most string compactifications, but is nevertheless well below the observational limit of $\Delta N_\text{eff} \lesssim 0.1$ (95\% confidence level)\cite{Goldstein:2026iuu}, which may be lowered by future experiments to 0.02 \cite{Baumann:2017gkg,NASAPICO:2019thw,Sehgal:2019ewc}. A second, lower peak occurs when $\epsilon$ reaches zero at large $N_a$ (cf.~Fig.~\ref{fig:effcurves}).
	
	The PBH mass can potentially be larger (see Fig.~\ref{fig:tevap}) depending on the number of axion species, but this can only increase $\Delta N_\text{eff}$ by an order of magnitude at most. Equation (\ref{DeltaNeff}) is already maximized at $a_0=0$, which is not substantially different from the cases shown in Fig.~\ref{fig:DeltaNeff}. Overall, we find $\Delta N_\text{eff} \lesssim 10^{-3}$ in the parametric range relevant for superradiant dark matter production.
	
	We note that $\Delta N_\text{eff}\propto N_a^{-1/2}$ for $N_a\gg 1$. While this may naively seem surprising, recall that for a large number of axion species these dominate the Hawking emission spectrum 
	($e_{T,a}\simeq e_T$), with Hawking emission being also the primary dark matter production channel. This implies that the ratio $(\rho_a/\rho_R)|_\text{evap}$ only depends on $N_a$ through $t_\text{evap}$, being suppressed for PBHs that evaporate earlier and hence for larger $N_a$.
	
	We have assumed so far that superradiance and Hawking emission together produce all of the dark matter. For completeness, if we assume that this makes up only a fraction $f<1$ of the dark matter observed at late times, $\mu \, Y_\text{DM} = f \times 0.43$ eV, then $\Delta N_\text{eff}$ will also be proportionally smaller, making this scenario even safer but further from observation.

	\subsection{PBH energy density}\label{sec:RhoPBH}
	
	In the large $N_a$ limit, superradiance is inefficient and the dark matter is produced essentially only by Hawking emission, while evaporation is dominated by emission of axions. This implies a large number density of PBHs to produce all of the dark matter, which could naively lead to an early PBH-dominated matter era. Similar to the above calculation of axion energy density, the energy density in PBHs is fixed in terms of the observed dark matter yield, and its maximum can be approximated as 
	\begin{align}
		\rho_\text{BH} &\lesssim s \, Y_\text{BH} \, M_0 \\
		&= s%{2\pi^2\over 25}g_\star \left({45\over 16\pi^3 g_\star} {\mpl^2\over t_\text{evap}^2}\right)^{3/4} 
		\, Y_\text{DM} {\left(1-\epsilon\right) \over N_\text{HE}} M_0
	\end{align}
	evaluated at evaporation. The largest $\rho_\text{BH}$ for the parameter choices considered above occurs with $M_0 = 5 \times 10^5$ kg, $a_0 = 0.02$, and $\alpha=0.004$, for which $\epsilon \to 0$ with $N_a \gtrsim 800$, and is tiny compared to the SM radiation, $\rho_\text{BH}/\rho_R  \lesssim 6.3 \times 10^{-5}$ at evaporation. In any case, we have only considered PBH masses such that they evaporate before BBN, and with many axion species they evaporate even earlier, so PBHs never dominate in the parametric range which we have considered.

	\section{Discussion}
	
	We have studied the purely gravitational production of heavy spin-0 dark matter through a combination of Hawking emission and superradiance by PBHs, showing that the latter's contribution is greatly enhanced by the existence of a number of light axion species $N_a\gtrsim 100$, which affect the dynamics of PBH evaporation. The PBHs are taken to have a (monochromatic) formation mass $M_0 \lesssim 10^6$ kg, so that they evaporate before BBN, low spins at or below the percent-level, and an abundance that leads to the presently measured dark matter density. The efficiency of superradiant dark matter production increases with the number of axion species $N_a$, up to a maximum number (dependent on the PBH mass and spin and the dark matter boson mass) above which evaporation into axions occurs so quickly that a superradiant cloud cannot form. Superradiance produces dark matter most efficiently with $100 \lesssim N_a \lesssim 10^6$, depending on the PBH formation mass and spin (see Figs.~\ref{fig:effcurves} and \ref{fig:1percent}), over a wide dark matter mass range, TeV$\lesssim \mu \lesssim$ PeV. 
	
	This means that for the $\mathcal{O}(100-10^5)$ light axion species expected in string compactifications, PBH superradiance may play a prominent role in dark matter production, alongside Hawking emission, even for slowly spinning PBHs. This is particularly relevant for dark matter searches, since superradiance is not only a purely gravitational mechanism, requiring no other interactions between dark matter and SM particles, but may also result in microscopic ``dark boson stars", i.e.~compact self-gravitating bound states of dark matter bosons with huge occupation numbers, albeit with sub-atomic radius. 
	
	This was originally conjectured in \cite{March-Russell:2022zll} and verified through numerical simulations in \cite{Neves:2025kxp}, which showed that a dark matter cloud first expands adiabatically as its PBH host evaporates up to close to $t_\text{evap}$, with the cloud's self-gravity becoming more and more significant as the PBH's mass decreases. Since PBH evaporation speeds up towards the end, $\dot{M}/M\sim M^{-3}$, eventually the cloud's wavefunction can no longer adjust adiabatically to the changing gravitational potential. Numerical simulations of the associated non-linear Schr\"odinger-Poisson system showed that, if the PBH mass is already $\lesssim 1/2$ the cloud's mass, then most of the dark matter bosons remain self-bound in a boson star, with only a small fraction becoming free particles. 
	
	A significant fraction of dark matter enclosed in these microscopic solitons suggests new avenues for dark matter detection: although the lack of (significant) non-gravitational interactions with e.g.~atomic nuclei or electrons could make detection of individual dark matter bosons impossible, dark boson stars may leave observable signals, as a result of the coherent enhancement of scattering cross-sections \cite{Hardy:2015boa}. This enhancement could be up to a factor $(N_{S}^\text{max})^2\sim 10^{50}$ assuming no significant mass loss in boson star formation and for a sufficiently compact boson star (compared to the inverse momentum exchange in the scattering process)\footnote{This is probably optimistic since rotating boson stars are unstable and decay into spherical ground states through scalar and gravitational radiation, although this should not change the order of magnitude of the cloud's mass \cite{Sanchis-Gual:2019ljs,DiGiovanni:2020ror,Dmitriev:2021utv}.}. Despite this huge enhancement, there is a price to pay: dark boson stars are relatively rare objects, with an expected flux of $\sim 3\times 10^{-4}(\mathrm{TeV}/\mu)(10^{25}/N_\text{S})\ \mathrm{km^{-2}yr^{-1}}$ in the solar neighbourhood, thus requiring large detection areas/exposure times. 
	
	Although a detailed phenomenological study of dark boson star interactions with SM particles is beyond the scope of this work, this nevertheless motivates investigating how boson star formation from PBH superradiance may rescue different ``nightmare scenarios'' for dark matter detection.
	
	We emphasize that the scenario we are considering could be truly a nightmare from the phenomenological perspective, although entirely plausible given what we currently know about dark matter: a heavy boson with no significant self-interactions or interactions with known particles stronger than gravity, carrying an approximately conserved global charge up to quantum-suppressed operators. Their production is purely gravitational, boosted only by the emission of light string axions, themselves behaving as dark radiation and giving only a small contribution to the number of relativistic species in the universe ($\Delta N_\text{eff}< 5 \times 10^{-4}$, the maximum being curiously attained for $N_a\sim \mathcal{O}(100)$ typical of most string compactifications).
	
	In future work we plan to extend our analysis to consider more realistic PBH mass and spin distributions, different types of (suppressed) couplings between dark matter and visible degrees of freedom and also the possibility that (at least some of) the string axions have non-negligible interactions with SM particles \cite{Dessert:2025yvk,Baryakhtar:2026oun}.

	\section{Acknowledgments}
	
	We thank Marco Calz\`a for useful discussion. This work was supported by national funds by FCT - Funda\c{c}\~ao para a Ci\^encia e Tecnologia, I.P., through the research project with DOI identifier 10.54499/UID/04564/2025, and by the project 10.54499/2024.00252.CERN funded by measure RE-C06-i06.m02 – ``Reinforcement of funding for International Partnerships in Science, Technology and Innovation'' of the Recovery and Resilience Plan - RRP, within the framework of the financing contract signed between the Recover Portugal Mission Structure (EMRP) and the Foundation for Science and Technology I.P. (FCT), as an intermediate beneficiary.


\begin{thebibliography}{0.5\textwidth}
		
%\cite{Carr:2020gox}
\bibitem{Carr:2020gox}
B.~Carr, K.~Kohri, Y.~Sendouda and J.~Yokoyama,
%``Constraints on primordial black holes,''
Rept. Prog. Phys. \textbf{84}, no.11, 116902 (2021)
%doi:10.1088/1361-6633/ac1e31
[arXiv:2002.12778 [astro-ph.CO]].
%1521 citations counted in INSPIRE as of 16 Jun 2026

%\cite{Bernal:2022oha}
\bibitem{Bernal:2022oha}
N.~Bernal, Y.~F.~Perez-Gonzalez and Y.~Xu,
%``Superradiant production of heavy dark matter from primordial black holes,''
Phys. Rev. D \textbf{106}, no.1, 015020 (2022)
%doi:10.1103/PhysRevD.106.015020
[arXiv:2205.11522 [hep-ph]].
%59 citations counted in INSPIRE as of 05 Jun 2026

%\cite{March-Russell:2022zll}
\bibitem{March-Russell:2022zll}
J.~March-Russell and J.~G.~Rosa,
%``Micro-Bose or Proca dark matter stars from black hole superradiance,''
Phys. Rev. D \textbf{113}, no.10, L101304 (2026)
%doi:10.1103/rwn4-8ywh
[arXiv:2205.15277 [gr-qc]].
%18 citations counted in INSPIRE as of 27 May 2026

%\cite{Jia:2025vqn}
\bibitem{Jia:2025vqn}
N.~Jia, S.~S.~Bao, C.~Zhang, H.~Zhang and X.~Zhang,
%``Superradiant dark matter production from primordial black holes: impact of multiple modes and gravitational wave emission,''
JHEP \textbf{09}, 195 (2025)
%doi:10.1007/JHEP09(2025)195
[arXiv:2504.18935 [astro-ph.CO]].
%8 citations counted in INSPIRE as of 24 Jun 2026

%\cite{Lennon:2017tqq}
\bibitem{Lennon:2017tqq}
O.~Lennon, J.~March-Russell, R.~Petrossian-Byrne and H.~Tillim,
%``Black Hole Genesis of Dark Matter,''
JCAP \textbf{04}, 009 (2018)
%doi:10.1088/1475-7516/2018/04/009
[arXiv:1712.07664 [hep-ph]].
%179 citations counted in INSPIRE as of 21 May 2026

%\cite{Allahverdi:2017sks}
\bibitem{Allahverdi:2017sks}
R.~Allahverdi, J.~Dent and J.~Osinski,
%``Nonthermal production of dark matter from primordial black holes,''
Phys. Rev. D \textbf{97}, no.5, 055013 (2018)
%doi:10.1103/PhysRevD.97.055013
[arXiv:1711.10511 [astro-ph.CO]].
%108 citations counted in INSPIRE as of 02 Jun 2026

%\cite{Hooper:2019gtx}
\bibitem{Hooper:2019gtx}
D.~Hooper, G.~Krnjaic and S.~D.~McDermott,
%``Dark Radiation and Superheavy Dark Matter from Black Hole Domination,''
JHEP \textbf{08}, 001 (2019)
%doi:10.1007/JHEP08(2019)001
[arXiv:1905.01301 [hep-ph]].
%231 citations counted in INSPIRE as of 22 May 2026

%\cite{Bernal:2020bjf}
\bibitem{Bernal:2020bjf}
N.~Bernal and {\'O}.~Zapata,
%``Dark Matter in the Time of Primordial Black Holes,''
JCAP \textbf{03}, 015 (2021)
%doi:10.1088/1475-7516/2021/03/015
[arXiv:2011.12306 [astro-ph.CO]].
%103 citations counted in INSPIRE as of 05 Jun 2026

%\cite{Bernal:2020ili}
\bibitem{Bernal:2020ili}
N.~Bernal and {\'O}.~Zapata,
%``Gravitational dark matter production: primordial black holes and UV freeze-in,''
Phys. Lett. B \textbf{815}, 136129 (2021)
%doi:10.1016/j.physletb.2021.136129
[arXiv:2011.02510 [hep-ph]].
%82 citations counted in INSPIRE as of 05 Jun 2026

%\cite{Masina:2020xhk}
\bibitem{Masina:2020xhk}
I.~Masina,
%``Dark matter and dark radiation from evaporating primordial black holes,''
Eur. Phys. J. Plus \textbf{135}, no.7, 552 (2020)
%doi:10.1140/epjp/s13360-020-00564-9
[arXiv:2004.04740 [hep-ph]].
%173 citations counted in INSPIRE as of 11 Jun 2026

%\cite{Baldes:2020nuv}
\bibitem{Baldes:2020nuv}
I.~Baldes, Q.~Decant, D.~C.~Hooper and L.~Lopez-Honorez,
%``Non-Cold Dark Matter from Primordial Black Hole Evaporation,''
JCAP \textbf{08}, 045 (2020)
%doi:10.1088/1475-7516/2020/08/045
[arXiv:2004.14773 [astro-ph.CO]].
%160 citations counted in INSPIRE as of 29 May 2026

%\cite{Gondolo:2020uqv}
\bibitem{Gondolo:2020uqv}
P.~Gondolo, P.~Sandick and B.~Shams Es Haghi,
%``Effects of primordial black holes on dark matter models,''
Phys. Rev. D \textbf{102}, no.9, 095018 (2020)
%doi:10.1103/PhysRevD.102.095018
[arXiv:2009.02424 [hep-ph]].
%148 citations counted in INSPIRE as of 05 Jun 2026

%\cite{Cheek:2021odj}
\bibitem{Cheek:2021odj}
A.~Cheek, L.~Heurtier, Y.~F.~Perez-Gonzalez and J.~Turner,
%``Primordial black hole evaporation and dark matter production. I. Solely Hawking radiation,''
Phys. Rev. D \textbf{105}, no.1, 015022 (2022)
%doi:10.1103/PhysRevD.105.015022
[arXiv:2107.00013 [hep-ph]].
%166 citations counted in INSPIRE as of 12 Jun 2026

%\cite{Cheek:2021cfe}
\bibitem{Cheek:2021cfe}
A.~Cheek, L.~Heurtier, Y.~F.~Perez-Gonzalez and J.~Turner,
%``Primordial black hole evaporation and dark matter production. II. Interplay with the freeze-in or freeze-out mechanism,''
Phys. Rev. D \textbf{105}, no.1, 015023 (2022)
%doi:10.1103/PhysRevD.105.015023
[arXiv:2107.00016 [hep-ph]].
%112 citations counted in INSPIRE as of 05 Jun 2026

%\cite{Chiba:2017rvs}
\bibitem{Chiba:2017rvs}
T.~Chiba and S.~Yokoyama,
%``Spin Distribution of Primordial Black Holes,''
PTEP \textbf{2017}, no.8, 083E01 (2017)
%doi:10.1093/ptep/ptx087
[arXiv:1704.06573 [gr-qc]].
%126 citations counted in INSPIRE as of 02 Jun 2026

%\cite{Mirbabayi:2019uph}
\bibitem{Mirbabayi:2019uph}
M.~Mirbabayi, A.~Gruzinov and J.~Nore{\~n}a,
%``Spin of Primordial Black Holes,''
JCAP \textbf{03}, 017 (2020)
%doi:10.1088/1475-7516/2020/03/017
[arXiv:1901.05963 [astro-ph.CO]].
%150 citations counted in INSPIRE as of 22 May 2026

%\cite{DeLuca:2019buf}
\bibitem{DeLuca:2019buf}
V.~De Luca, V.~Desjacques, G.~Franciolini, A.~Malhotra and A.~Riotto,
%``The initial spin probability distribution of primordial black holes,''
JCAP \textbf{05}, 018 (2019)
%doi:10.1088/1475-7516/2019/05/018
[arXiv:1903.01179 [astro-ph.CO]].
%200 citations counted in INSPIRE as of 10 Jun 2026

%\cite{Harada:2017fjm}
\bibitem{Harada:2017fjm}
T.~Harada, C.~M.~Yoo, K.~Kohri and K.~I.~Nakao,
%``Spins of primordial black holes formed in the matter-dominated phase of the Universe,''
Phys. Rev. D \textbf{96}, no.8, 083517 (2017)
[erratum: Phys. Rev. D \textbf{99}, no.6, 069904 (2019)]
%doi:10.1103/PhysRevD.96.083517
[arXiv:1707.03595 [gr-qc]].
%215 citations counted in INSPIRE as of 12 Jun 2026

%\cite{Flores:2021tmc}
\bibitem{Flores:2021tmc}
M.~M.~Flores and A.~Kusenko,
%``Spins of primordial black holes formed in different cosmological scenarios,''
Phys. Rev. D \textbf{104}, no.6, 063008 (2021)
%doi:10.1103/PhysRevD.104.063008
[arXiv:2106.03237 [astro-ph.CO]].
%44 citations counted in INSPIRE as of 28 May 2026

%\cite{Saito:2023fpt}
\bibitem{Saito:2023fpt}
D.~Saito, T.~Harada, Y.~Koga and C.~M.~Yoo,
%``Spins of primordial black holes formed with a soft equation of state,''
JCAP \textbf{07}, 030 (2023)
%doi:10.1088/1475-7516/2023/07/030
[arXiv:2305.13830 [gr-qc]].
%15 citations counted in INSPIRE as of 02 Jun 2026

%\cite{Saito:2024hlj}
\bibitem{Saito:2024hlj}
D.~Saito, T.~Harada, Y.~Koga and C.~M.~Yoo,
%``Revisiting spins of primordial black holes in a matter-dominated era based on peak theory,''
JCAP \textbf{11}, 064 (2024)
%doi:10.1088/1475-7516/2024/11/064
[arXiv:2409.00435 [gr-qc]].
%9 citations counted in INSPIRE as of 02 Jun 2026

%\cite{Neves:2025kxp}
\bibitem{Neves:2025kxp}
D.~Neves and J.~G.~Rosa,
%``From scalar clouds around evaporating black holes to boson star,''
[arXiv:2512.03155 [gr-qc]].
%1 citations counted in INSPIRE as of 21 May 2026

%\cite{Hardy:2015boa}
\bibitem{Hardy:2015boa}
E.~Hardy, R.~Lasenby, J.~March-Russell and S.~M.~West,
%``Signatures of Large Composite Dark Matter States,''
JHEP \textbf{07}, 133 (2015)
%doi:10.1007/JHEP07(2015)133
[arXiv:1504.05419 [hep-ph]].
%75 citations counted in INSPIRE as of 28 Apr 2026

%\cite{Fallon:2025lvn}
\bibitem{Fallon:2025lvn}
S.~V.~Fallon, J.~Halverson, L.~McAllister and Y.~Zhu,
%``F-theory Axiverse,''
[arXiv:2511.20458 [hep-th]].
%15 citations counted in INSPIRE as of 15 Jun 2026

%\cite{Svrcek:2006yi}
\bibitem{Svrcek:2006yi}
P.~Svrcek and E.~Witten,
%``Axions In String Theory,''
JHEP \textbf{06}, 051 (2006)
%doi:10.1088/1126-6708/2006/06/051
[arXiv:hep-th/0605206 [hep-th]].
%2046 citations counted in INSPIRE as of 17 Jun 2026

%\cite{Arvanitaki:2009fg}
\bibitem{Arvanitaki:2009fg}
A.~Arvanitaki, S.~Dimopoulos, S.~Dubovsky, N.~Kaloper and J.~March-Russell,
%``String Axiverse,''
Phys. Rev. D \textbf{81}, 123530 (2010)
%doi:10.1103/PhysRevD.81.123530
[arXiv:0905.4720 [hep-th]].
%2451 citations counted in INSPIRE as of 17 Jun 2026

%\cite{Arvanitaki:2010sy}
\bibitem{Arvanitaki:2010sy}
A.~Arvanitaki and S.~Dubovsky,
%``Exploring the String Axiverse with Precision Black Hole Physics,''
Phys. Rev. D \textbf{83}, 044026 (2011)
%doi:10.1103/PhysRevD.83.044026
[arXiv:1004.3558 [hep-th]].
%792 citations counted in INSPIRE as of 12 Jun 2026

%\cite{Gendler:2023kjt}
\bibitem{Gendler:2023kjt}
N.~Gendler, D.~J.~E.~Marsh, L.~McAllister and J.~Moritz,
%``Glimmers from the axiverse,''
JCAP \textbf{09}, 071 (2024)
%doi:10.1088/1475-7516/2024/09/071
[arXiv:2309.13145 [hep-th]].
%98 citations counted in INSPIRE as of 17 Jun 2026

%\cite{Calza:2021czr}
\bibitem{Calza:2021czr}
M.~Calz{\`a}, J.~March-Russell and J.~G.~Rosa,
%``Evaporating Primordial Black Holes, the String Axiverse, and Hot Dark Radiation,''
Phys. Rev. Lett. \textbf{133}, no.26, 261003 (2024)
%doi:10.1103/PhysRevLett.133.261003
[arXiv:2110.13602 [astro-ph.CO]].
%56 citations counted in INSPIRE as of 05 Jun 2026

%\cite{Calza:2023rjt}
\bibitem{Calza:2023rjt}
M.~Calz{\`a}, J.~G.~Rosa and F.~Serrano,
%``Primordial black hole superradiance and evaporation in the string axiverse,''
JHEP \textbf{05}, 140 (2024)
%doi:10.1007/JHEP05(2024)140
[arXiv:2306.09430 [hep-ph]].
%43 citations counted in INSPIRE as of 05 Jun 2026

%\cite{Marsh:2015xka}
\bibitem{Marsh:2015xka}
D.~J.~E.~Marsh,
%``Axion Cosmology,''
Phys. Rept. \textbf{643}, 1-79 (2016)
%doi:10.1016/j.physrep.2016.06.005
[arXiv:1510.07633 [astro-ph.CO]].
%2050 citations counted in INSPIRE as of 17 Jun 2026

%\cite{OHare:2024nmr}
\bibitem{OHare:2024nmr}
C.~A.~J.~O'Hare,
%``Cosmology of axion dark matter,''
PoS \textbf{COSMICWISPers}, 040 (2024)
%doi:10.22323/1.454.0040
[arXiv:2403.17697 [hep-ph]].
%195 citations counted in INSPIRE as of 17 Jun 2026

%\cite{Chen:2014jwq}
\bibitem{Chen:2014jwq}
P.~Chen, Y.~C.~Ong and D.~h.~Yeom,
%``Black Hole Remnants and the Information Loss Paradox,''
Phys. Rept. \textbf{603}, 1-45 (2015)
%doi:10.1016/j.physrep.2015.10.007
[arXiv:1412.8366 [gr-qc]].
%362 citations counted in INSPIRE as of 12 Jun 2026

%\cite{Baryakhtar:2020gao}
\bibitem{Baryakhtar:2020gao}
M.~Baryakhtar, M.~Galanis, R.~Lasenby and O.~Simon,
%``Black hole superradiance of self-interacting scalar fields,''
Phys. Rev. D \textbf{103}, no.9, 095019 (2021)
%doi:10.1103/PhysRevD.103.095019
[arXiv:2011.11646 [hep-ph]].
%295 citations counted in INSPIRE as of 05 Jun 2026

%\cite{Branco:2023frw}
\bibitem{Branco:2023frw}
N.~P.~Branco, R.~Z.~Ferreira and J.~G.~Rosa,
%``Superradiant axion clouds around asteroid-mass primordial black holes,''
JCAP \textbf{04}, 003 (2023)
%doi:10.1088/1475-7516/2023/04/003
[arXiv:2301.01780 [hep-ph]].
%37 citations counted in INSPIRE as of 05 Jun 2026

%\cite{Xie:2025npy}
\bibitem{Xie:2025npy}
N.~Xie and F.~P.~Huang,
%``Self-interaction effects on the Kerr black hole superradiance and their observational implications,''
Phys. Rev. D \textbf{112}, no.5, 055028 (2025)
%doi:10.1103/xmhn-cpv4
[arXiv:2503.10347 [hep-ph]].
%12 citations counted in INSPIRE as of 02 Jun 2026

%\cite{Hawking:1975vcx}
\bibitem{Hawking:1975vcx}
S.~W.~Hawking,
%``Particle Creation by Black Holes,''
Commun. Math. Phys. \textbf{43}, 199-220 (1975)
[erratum: Commun. Math. Phys. \textbf{46}, 206 (1976)]
%doi:10.1007/BF02345020
%13587 citations counted in INSPIRE as of 17 Jun 2026

%\cite{Hartle:1976tp}
\bibitem{Hartle:1976tp}
J.~B.~Hartle and S.~W.~Hawking,
%``Path Integral Derivation of Black Hole Radiance,''
Phys. Rev. D \textbf{13}, 2188-2203 (1976)
%doi:10.1103/PhysRevD.13.2188
%1264 citations counted in INSPIRE as of 15 Jun 2026

%\cite{Starobinskii:1973vzb}
\bibitem{Starobinskii:1973vzb}
A.~A.~Starobinskii,
%``Amplification of waves during reflection from a rotating ''black hole'',''
Sov. Phys. JETP \textbf{37}, no.1, 28-32 (1973)
%583 citations counted in INSPIRE as of 10 Jun 2026

%\cite{Cardoso:2004nk}
\bibitem{Cardoso:2004nk}
V.~Cardoso, O.~J.~C.~Dias, J.~P.~S.~Lemos and S.~Yoshida,
%``The Black hole bomb and superradiant instabilities,''
Phys. Rev. D \textbf{70}, 044039 (2004)
[erratum: Phys. Rev. D \textbf{70}, 049903 (2004)]
%doi:10.1103/PhysRevD.70.049903
[arXiv:hep-th/0404096 [hep-th]].
%442 citations counted in INSPIRE as of 08 Jun 2026

%\cite{Pani:2012bp}
\bibitem{Pani:2012bp}
P.~Pani, V.~Cardoso, L.~Gualtieri, E.~Berti and A.~Ishibashi,
%``Perturbations of slowly rotating black holes: massive vector fields in the Kerr metric,''
Phys. Rev. D \textbf{86}, 104017 (2012)
%doi:10.1103/PhysRevD.86.104017
[arXiv:1209.0773 [gr-qc]].
%277 citations counted in INSPIRE as of 09 Jun 2026

%\cite{Rosa:2012uz}
\bibitem{Rosa:2012uz}
J.~G.~Rosa,
%``Boosted black string bombs,''
JHEP \textbf{02}, 014 (2013)
%doi:10.1007/JHEP02(2013)014
[arXiv:1209.4211 [hep-th]].
%47 citations counted in INSPIRE as of 05 Jun 2026

%\cite{Pani:2012vp}
\bibitem{Pani:2012vp}
P.~Pani, V.~Cardoso, L.~Gualtieri, E.~Berti and A.~Ishibashi,
%``Black hole bombs and photon mass bounds,''
Phys. Rev. Lett. \textbf{109}, 131102 (2012)
%doi:10.1103/PhysRevLett.109.131102
[arXiv:1209.0465 [gr-qc]].
%300 citations counted in INSPIRE as of 14 Apr 2026

%\cite{Witek:2012tr}
\bibitem{Witek:2012tr}
H.~Witek, V.~Cardoso, A.~Ishibashi and U.~Sperhake,
%``Superradiant instabilities in astrophysical systems,''
Phys. Rev. D \textbf{87}, no.4, 043513 (2013)
%doi:10.1103/PhysRevD.87.043513
[arXiv:1212.0551 [gr-qc]].
%268 citations counted in INSPIRE as of 10 Apr 2026

%\cite{Brito:2014wla}
\bibitem{Brito:2014wla}
R.~Brito, V.~Cardoso and P.~Pani,
%``Black holes as particle detectors: evolution of superradiant instabilities,''
Class. Quant. Grav. \textbf{32}, no.13, 134001 (2015)
%doi:10.1088/0264-9381/32/13/134001
[arXiv:1411.0686 [gr-qc]].
%377 citations counted in INSPIRE as of 15 May 2026

%\cite{Arvanitaki:2014wva}
\bibitem{Arvanitaki:2014wva}
A.~Arvanitaki, M.~Baryakhtar and X.~Huang,
%``Discovering the QCD Axion with Black Holes and Gravitational Waves,''
Phys. Rev. D \textbf{91}, no.8, 084011 (2015)
%doi:10.1103/PhysRevD.91.084011
[arXiv:1411.2263 [hep-ph]].
%650 citations counted in INSPIRE as of 12 Jun 2026

%\cite{Brito:2015oca}
\bibitem{Brito:2015oca}
R.~Brito, V.~Cardoso and P.~Pani,
%``Superradiance: New Frontiers in Black Hole
Physics,''
Lect. Notes Phys. \textbf{906}, pp.1-237 (2015)
2020,
ISBN 978-3-319-18999-4, 978-3-319-19000-6, 978-3-030-46621-3, 978-3-030-46622-0
%doi:10.1007/978-3-319-19000-6
[arXiv:1501.06570 [gr-qc]].
%1031 citations counted in INSPIRE as of 17 Jun 2026

%\cite{Rosa:2017ury}
\bibitem{Rosa:2017ury}
J.~G.~Rosa and T.~W.~Kephart,
%``Stimulated Axion Decay in Superradiant Clouds around Primordial Black Holes,''
Phys. Rev. Lett. \textbf{120}, no.23, 231102 (2018)
%doi:10.1103/PhysRevLett.120.231102
[arXiv:1709.06581 [gr-qc]].
%143 citations counted in INSPIRE as of 10 Jun 2026

%\cite{Cardoso:2018tly}
\bibitem{Cardoso:2018tly}
V.~Cardoso, {\'O}.~J.~C.~Dias, G.~S.~Hartnett, M.~Middleton, P.~Pani and J.~E.~Santos,
%``Constraining the mass of dark photons and axion-like particles through black-hole superradiance,''
JCAP \textbf{03}, 043 (2018)
%doi:10.1088/1475-7516/2018/03/043
[arXiv:1801.01420 [gr-qc]].
%290 citations counted in INSPIRE as of 10 Jun 2026

%\cite{Baumann:2018vus}
\bibitem{Baumann:2018vus}
D.~Baumann, H.~S.~Chia and R.~A.~Porto,
%``Probing Ultralight Bosons with Binary Black Holes,''
Phys. Rev. D \textbf{99}, no.4, 044001 (2019)
%doi:10.1103/PhysRevD.99.044001
[arXiv:1804.03208 [gr-qc]].
%350 citations counted in INSPIRE as of 10 Jun 2026

%\cite{Baumann:2019eav}
\bibitem{Baumann:2019eav}
D.~Baumann, H.~S.~Chia, J.~Stout and L.~ter Haar,
%``The Spectra of Gravitational Atoms,''
JCAP \textbf{12}, 006 (2019)
%doi:10.1088/1475-7516/2019/12/006
[arXiv:1908.10370 [gr-qc]].
%231 citations counted in INSPIRE as of 10 Jun 2026

%\cite{Page:1976ki}
\bibitem{Page:1976ki}
D.~N.~Page,
%``Particle Emission Rates from a Black Hole. 2. Massless Particles from a Rotating Hole,''
Phys. Rev. D \textbf{14}, 3260-3273 (1976)
%doi:10.1103/PhysRevD.14.3260
%658 citations counted in INSPIRE as of 15 Jun 2026

%\cite{Chambers:1997ai}
\bibitem{Chambers:1997ai}
C.~M.~Chambers, W.~A.~Hiscock and B.~Taylor,
%``Spinning down a black hole with scalar fields,''
Phys. Rev. Lett. \textbf{78}, 3249-3251 (1997)
%doi:10.1103/PhysRevLett.78.3249
[arXiv:gr-qc/9703018 [gr-qc]].
%37 citations counted in INSPIRE as of 01 Apr 2026

%\cite{Taylor:1998dk}
\bibitem{Taylor:1998dk}
B.~E.~Taylor, C.~M.~Chambers and W.~A.~Hiscock,
%``Evaporation of a Kerr black hole by emission of scalar and higher spin particles,''
Phys. Rev. D \textbf{58}, 044012 (1998)
%doi:10.1103/PhysRevD.58.044012
[arXiv:gr-qc/9801044 [gr-qc]].
%63 citations counted in INSPIRE as of 12 Jun 2026

%\cite{Kofman:1982gu}
\bibitem{Kofman:1982gu}
L.~A.~Kofman,
%``BOUND STATES IN QUANTUM EVAPORATION OF BLACK HOLES,''
Phys. Lett. A \textbf{87}, 281-284 (1982)
%doi:10.1016/0375-9601(82)90696-X
%9 citations counted in INSPIRE as of 04 Dec 2025

%\cite{Fu:2025ztk}
\bibitem{Fu:2025ztk}
L.~Fu, H.~Omiya, T.~Tanaka, X.~Tong, Y.~Wang and H.~Y.~Zhu,
%``Quantum Treatment of Black Hole Superradiance,''
[arXiv:2512.06790 [gr-qc]].
%3 citations counted in INSPIRE as of 13 May 2026

%\cite{Planck:2018vyg}
\bibitem{Planck:2018vyg}
N.~Aghanim \textit{et al.} [Planck],
%``Planck 2018 results. VI. Cosmological parameters,''
Astron. Astrophys. \textbf{641}, A6 (2020)
[erratum: Astron. Astrophys. \textbf{652}, C4 (2021)]
%doi:10.1051/0004-6361/201833910
[arXiv:1807.06209 [astro-ph.CO]].
%22067 citations counted in INSPIRE as of 17 Jun 2026

%\cite{Hooper:2020evu}
\bibitem{Hooper:2020evu}
D.~Hooper, G.~Krnjaic, J.~March-Russell, S.~D.~McDermott and R.~Petrossian-Byrne,
%``Hot Gravitons and Gravitational Waves From Kerr Black Holes in the Early Universe,''
[arXiv:2004.00618 [astro-ph.CO]].
%117 citations counted in INSPIRE as of 21 May 2026

%\cite{Arbey:2021ysg}
\bibitem{Arbey:2021ysg}
A.~Arbey, J.~Auffinger, P.~Sandick, B.~Shams Es Haghi and K.~Sinha,
%``Precision calculation of dark radiation from spinning primordial black holes and early matter-dominated eras,''
Phys. Rev. D \textbf{103}, no.12, 123549 (2021)
%doi:10.1103/PhysRevD.103.123549
[arXiv:2104.04051 [astro-ph.CO]].
%88 citations counted in INSPIRE as of 05 Jun 2026

%\cite{Franciolini:2026fdv}
\bibitem{Franciolini:2026fdv}
G.~Franciolini and D.~Racco,
%``Isocurvature constraints on dark matter from evaporated primordial black holes,''
Phys. Rev. D \textbf{113}, no.12, 123505 (2026)
%doi:10.1103/xjvh-tpjw
[arXiv:2603.02322 [astro-ph.CO]].
%6 citations counted in INSPIRE as of 04 Jun 2026

%\cite{Jia:2026psi}
\bibitem{Jia:2026psi}
N.~Jia, C.~Zhang and X.~Zhang,
%``Dark radiation from Kerr primordial black holes: the role of superradiance,''
[arXiv:2603.29790 [astro-ph.CO]].
%2 citations counted in INSPIRE as of 17 Jun 2026

%\cite{Goldstein:2026iuu}
\bibitem{Goldstein:2026iuu}
S.~Goldstein and J.~C.~Hill,
%``A 2{\%} determination of $N_{\rm eff}$ from primordial element abundance, cosmic microwave background, and baryon acoustic oscillation measurements,''
[arXiv:2603.13226 [astro-ph.CO]].
%7 citations counted in INSPIRE as of 18 Jun 2026

%\cite{Baumann:2017gkg}
\bibitem{Baumann:2017gkg}
D.~Baumann, D.~Green and B.~Wallisch,
%``Searching for light relics with large-scale structure,''
JCAP \textbf{08}, 029 (2018)
%doi:10.1088/1475-7516/2018/08/029
[arXiv:1712.08067 [astro-ph.CO]].
%130 citations counted in INSPIRE as of 06 May 2026

%\cite{NASAPICO:2019thw}
\bibitem{NASAPICO:2019thw}
S.~Hanany \textit{et al.} [NASA PICO],
%``PICO: Probe of Inflation and Cosmic Origins,''
[arXiv:1902.10541 [astro-ph.IM]].
%344 citations counted in INSPIRE as of 27 May 2026

%\cite{Sehgal:2019ewc}
\bibitem{Sehgal:2019ewc}
N.~Sehgal, S.~Aiola, Y.~Akrami, K.~Basu, M.~Boylan-Kolchin, S.~Bryan, S.~Clesse, F.~Y.~Cyr-Racine, L.~Di Mascolo and S.~Dicker, \textit{et al.}
%``CMB-HD: An Ultra-Deep, High-Resolution Millimeter-Wave Survey Over Half the Sky,''
Bull. Am. Astron. Soc. \textbf{51}, no.7, 1-23 (2019)
[arXiv:1906.10134 [astro-ph.CO]].
%173 citations counted in INSPIRE as of 09 Jun 2026

%\cite{Sanchis-Gual:2019ljs}
\bibitem{Sanchis-Gual:2019ljs}
N.~Sanchis-Gual, F.~Di Giovanni, M.~Zilh{\~a}o, C.~Herdeiro, P.~Cerd{\'a}-Dur{\'a}n, J.~A.~Font and E.~Radu,
%``Nonlinear Dynamics of Spinning Bosonic Stars: Formation and Stability,''
Phys. Rev. Lett. \textbf{123}, no.22, 221101 (2019)
%doi:10.1103/PhysRevLett.123.221101
[arXiv:1907.12565 [gr-qc]].
%149 citations counted in INSPIRE as of 15 Jun 2026

%\cite{DiGiovanni:2020ror}
\bibitem{DiGiovanni:2020ror}
F.~Di Giovanni, N.~Sanchis-Gual, P.~Cerd{\'a}-Dur{\'a}n, M.~Zilh{\~a}o, C.~Herdeiro, J.~A.~Font and E.~Radu,
%``Dynamical bar-mode instability in spinning bosonic stars,''
Phys. Rev. D \textbf{102}, no.12, 124009 (2020)
%doi:10.1103/PhysRevD.102.124009
[arXiv:2010.05845 [gr-qc]].
%73 citations counted in INSPIRE as of 12 May 2026

%\cite{Dmitriev:2021utv}
\bibitem{Dmitriev:2021utv}
A.~S.~Dmitriev, D.~G.~Levkov, A.~G.~Panin, E.~K.~Pushnaya and I.~I.~Tkachev,
%``Instability of rotating Bose stars,''
Phys. Rev. D \textbf{104}, no.2, 023504 (2021)
%doi:10.1103/PhysRevD.104.023504
[arXiv:2104.00962 [gr-qc]].
%65 citations counted in INSPIRE as of 15 Jun 2026

%\cite{Dessert:2025yvk}
\bibitem{Dessert:2025yvk}
C.~Dessert, S.~Kumar and J.~T.~Ruderman,
%``Freezing-in the Axiverse,''
[arXiv:2511.09631 [hep-ph]].
%5 citations counted in INSPIRE as of 15 Jun 2026

%\cite{Baryakhtar:2026oun}
\bibitem{Baryakhtar:2026oun}
M.~Baryakhtar, D.~Cyncynates and E.~Henry,
%``Axiverse Lampposts,''
[arXiv:2602.23424 [hep-ph]].
%5 citations counted in INSPIRE as of 15 Jun 2026
		
		
		
	\end{thebibliography}
\end{document}